\documentclass[aps,pre,10pt,twocolumn,superscriptaddress]{revtex4-2}
\usepackage{graphicx}
\usepackage{float}
\usepackage{epstopdf}
\usepackage{verbatim}
\usepackage{xcolor}
\usepackage{amsmath}
\usepackage{amssymb}
\usepackage[utf8]{inputenc}
\usepackage{graphicx,float,hyperref}
\usepackage{natbib}
\DeclareMathOperator{\sech}{sech}
\usepackage{calligra}  
\usepackage{graphicx,latexsym} 
\begin{document}
	
	
	\title{{\large {\Large Asymmetric Quantum Harmonic Otto Engine Under Hot Squeezed Thermal Reservoir}}}{\normalsize {\large}}
	
	\author{Monika}
	\email{monika.ph.23@nitj.ac.in}
	\affiliation{Department of Physics, Dr B R Ambedkar National Institute of Technology Jalandhar, Punjab-144008, India}

	\author{Kirandeep Kaur}
	\email{kiran.nitj@gmail.com}
	\affiliation{Department of Physical Sciences, Indian Institute of Science Education and Research Mohali, Sector-81, Manauli, Punjab-140306, India}
	
	\author{Varinder Singh}
	\email{varinderkias@kias.re.kr}
	\affiliation{School of Physics, Korea Institute for Advanced Study, Seoul 02455, Korea}
	
	\author{Shishram Rebari}
	\email{rebaris@nitj.ac.in}
	
	\affiliation{Department of Physics, Dr B R Ambedkar National Institute of Technology Jalandhar, Punjab-144008, India}
	
	\begin{abstract}
		We study a quantum harmonic Otto engine under a hot squeezed thermal reservoir with asymmetry  between  the two adiabatic branches introduced by considering different speeds of the driving protocols. In the first configuration, the driving protocol for the expansion  stroke  is  sudden-switch in nature and compression stroke is driven adiabatically, while the second configuration deals with the converse situation. In both cases, we obtain analytic expressions for the upper bound on efficiency and efficiency at  optimal  work output, which reveals a significant difference between the   two configurations.  Additionally, we  find that the maximum achievable efficiency in sudden expansion case is 1/2 only while it approaches unity for the sudden compression stroke. Further, we study the effect of increasing degree of squeezing on the efficiency and work output of the engine and indicate the optimal operational regime for both configurations under consideration.   Finally, by studying the full phase-diagram of the Otto cycle we observe that the operational region of the engine mode grows with  increasing  squeezing at the expense of refrigeration regime.
	\end{abstract}
	
	\maketitle 
	
	\section{Introduction}
	
	Heat engines are the most frequently used practical thermal devices that covert heat from a thermal reservoir to useful work\cite{cengel2011thermodynamics,borgnakke2020fundamentals,kaur2021unified,kaur2024optimization,singh2018low}. Efficiency of all heat engines working with thermal resources is bounded by Carnot efficiency, $\eta_c=1-T_c/T_h$, where $T_h$  $(T_c)$  is the temperature of the hot (cold) reservoir. However, quantum heat engines exploiting quantum resources, such as  squeezed reservoirs, may surpass Carnot efficiency without violating the second law of thermodynamics \cite{klaers2017squeezed,manzano2016entropy,kosloff2017quantum,huang2012effects,manzano2018squeezed,de2020universal,wang2019finite,singh2019three,huang2012effects}. Moreover, with the help of squeezing, the systems away from the thermal equilibrium can be explored as it provides better control over quantum states\cite{schnabel2017squeezed,fearn1988representations,marian1993squeezed}. 
	
	In recent years, the quantum Otto cycle has been widely studied by using different working materials\cite{son2021monitoring,piccitto2022ising,myers2021quantum,kaur2024performance,shastri2022optimization,harunari2021maximal,singh2023asymmetric}. Due to its amenability to analytical calculations, one of the most popular choice of working material in quantum heat engines is a time-dependent harmonic oscillator. The study of these quantum heat engines, along the squeezed thermal reservoirs, has gained significant attention in quantum thermodynamics studies\cite{singh2020performance,zhang2020optimization,wang2019finite,huang2012effects}. Squeezing facilitates the work extraction from a single squeezed reservoir, while a similar task is impossible with a standard thermal reservoir\cite{manzano2016entropy}. Recently, Ro{\ss}nagel $et$ $al.$ have studied a quantum Otto engine coupled to a hot squeezed thermal reservoir and claimed the efficiency at maximum work to surpass the Carnot bound\cite{rossnagel2014nanoscale}.

	Our work explores the thermodynamic performance of an asymmetric  quantum Otto cycle \cite{singh2022unified} with time dependent harmonic oscillator as the working fluid. We introduce asymmetry between two work strokes  by  driving them with different speeds, based on which we discuss two different configurations: one in which  expansion stroke is driven by sudden-switch protocol and compression strokes is adiabatic (slowly driven) in  nature, while the other configuration deals with the converse situation. Further, to investigate the advantages of squeezing, we couple the Otto engine to a hot squeezed reservoir while the cold reservoir is still purely thermal. We find analytical expressions of the upper bound on the efficiency and efficiency at maximum work output for both the cases mentioned above. Additionally, we will discuss the complete phase diagram of the Otto cycle under the squeezing effect.

	The  structured of the paper is   outlined below. Sec. II briefly discusses the model of the quantum Otto engine coupled to a hot squeezed reservoir. In Sec. III, we derive analytical expressions for the upper bound on efficiency and efficiency at maximum work. In Sec. III A, we will discuss engine with the sudden compression stroke, while  Sec. III B is devoted to a detailed study of the engine experiencing sudden expansion stroke. In Sec. IV, we will discuss the full phase diagram of the Otto cycle under the influence of squeezing. We will sum up our findings in Sec. V.

	\section{Quantum Otto Engine with hot squeezed reservoir}
\begin{figure}
		\includegraphics[width=8.6cm,height=6cm]{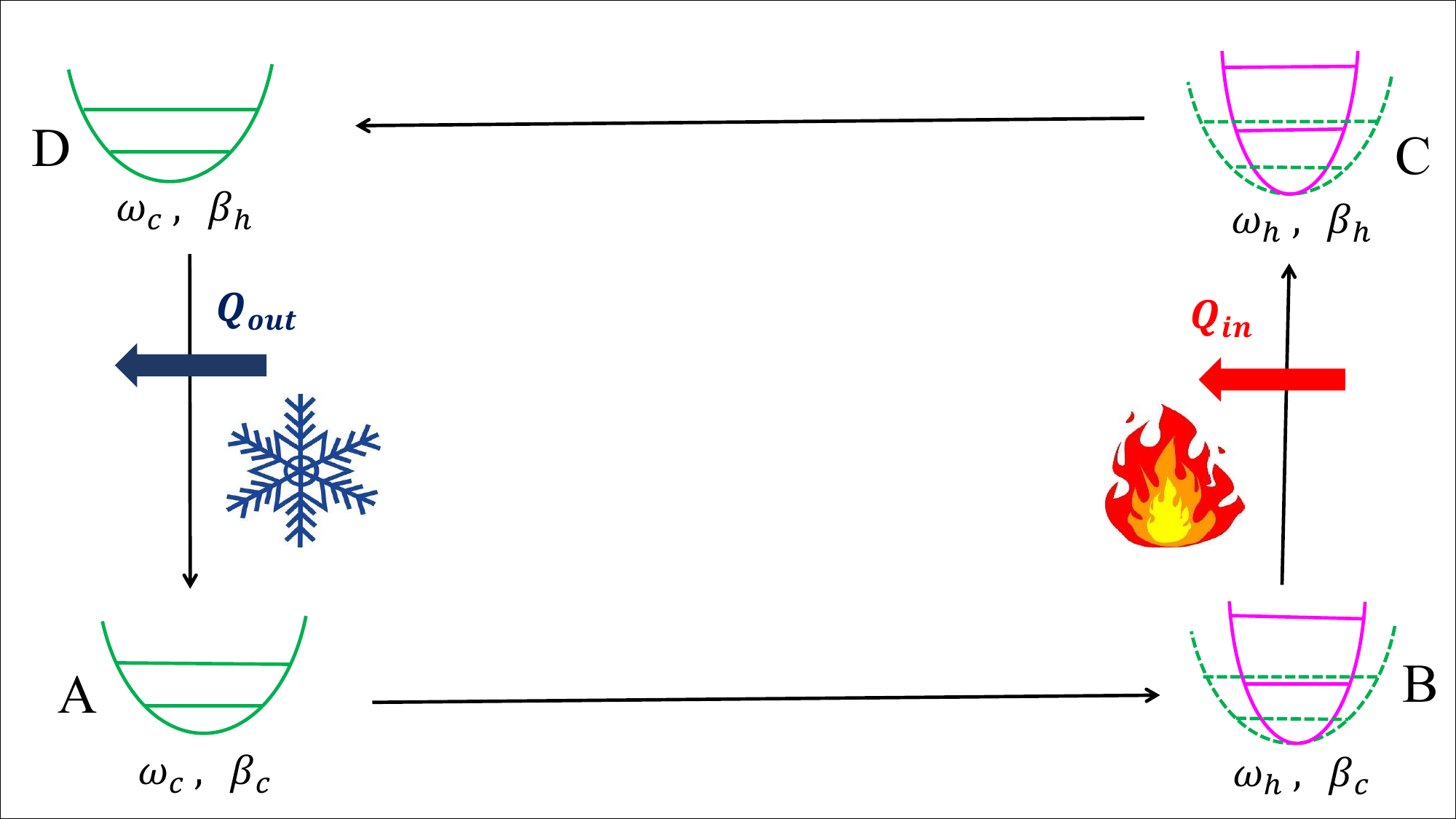}
		\caption{Diagrammatic representation of a quantum Otto cycle using a time-dependent harmonic oscillator as a working substance.}
		\label{fig:1}
	\end{figure}

	The Otto cycle consists of four processes, which are briefly discussed as follows (see Fig.\ref{fig:1})\cite{singh2020performance,kaur2024performance,nautiyal2024out,nautiyal2024finite}:
	
	(1)Adiabatic Compression ($A\rightarrow B$): Initially, the system is at inverse temperature $\beta_{c}$ = $(k_B T_{c})^{-1}$. Then, by using some external control, the oscillator's frequency is changed from $\omega_{c}$ to $\omega_{h}$ and the system states from A ($\omega_{c}$, $\beta_{c}$) to B ($\omega_{h}$, $\beta_{c}$, $\lambda_{AB}$). Work is added to the the system during this stage.   
	(2) Hot Isochore ($B\rightarrow C$): Now, place the system in contact with a hot squeezed thermal reservoir having inverse temperature $\beta_{h}$ = $(k_B T_{h})^{-1}$. Frequency remains constant during the isochoric process, and no work is done on (by) the system. Heat ($Q_{h}$) is added to the system and the system relaxes to the state C($\omega_{h}$, $\beta_{h}$, $r$).
	(3) Adiabatic Expansion ($C\rightarrow D$): At this stage, the system is detached from the hot squeezed reservoir, the frequency of the system is unitarily changed back to its initial value $\omega_{c}$, and the system gets tuned to the state D ($\omega_{c}$, $\beta_{h}$, $\lambda_{CD}$). In this stage, the work is extracted from the system. 
	(4) Cold Isochore ($D\rightarrow A$): Finally,   the system is put in thermal contact with   a cold thermal reservoir having inverse temperature $\beta_{c}$ ($\beta_{c}$ $>$ $\beta_{h}$). Heat ($Q_{c}$) releases from the system to the cold reservoir, and the system returns to its initial state A ($\omega_{c}$, $\beta_{c}$). Frequency $\omega_{c}$ remains constant during this process.
	
	During the above four processes, the average energies $\ <H>_{i}$ ($i$ = A, B, C, D) of the oscillator are as follows($\ k_{B}$ = $\hbar = 1$)\cite{husimi1953miscellanea}:
	\begin{align}
		\langle H \rangle_{A} &= \frac{\omega_{c}}{2} \coth \left(\frac{\beta_{c} \omega_{c}}{2}\right),\label{eq:1} \\
		\langle H \rangle_{B} &= \frac{\omega_{h}}{2}\lambda_{AB} \coth \left(\frac{\beta_{c} \omega_{c}}{2}\right), \label{eq:2}\\
		\langle H \rangle_{C} &= \frac{\omega_{h}}{2} \coth \left(\frac{\beta_{h} \omega_{h}}{2}\right)\cosh(2r), \label{eq:3}\\
		\langle H \rangle_{D} &= \frac{\omega_{c}}{2}\lambda_{CD} \coth \left(\frac{\beta_{h} \omega_{h}}{2}\right)\cosh(2r), \label{eq:4}
	\end{align}
	where the factor cosh(2$r$) is due to the squeezing of hot thermal reservoir with $r$ as a squeezed parameter; $ \lambda_{i}$ ($i$ = AB, CD) is the adiabaticity parameter that measures the degree of adiabaticity. In general, we have $\lambda_i \geq 1$ and is dimensionless. Here, we are interested in two extreme cases: adiabatic (slowly driving) and sudden-switch driving. For adiabatic processes, $\lambda_i$ = 1 and $\lambda_i$ = $(\omega^2_{c}+\omega^2_{h})/2\omega_{c} \omega_{h}$ $>$ 1 for sudden-switch cases. 
	
	The expressions for input heat and rejected heat during the hot and cold isochores, respectively, are given by
	\begin{align}
		\langle Q \rangle_{h} &= \langle H \rangle_{C} - \langle H \rangle_{B} \nonumber \\
		&= \frac{\omega_{h}}{2} \left[\coth \left(\frac{\beta_{h} \omega_{h}}{2}\right)\cosh(2r) - \lambda_{AB} \coth \left(\frac{\beta_{c} \omega_{c}}{2}\right)\right]. \label{eq:5} 
	\end{align} 
	\begin{align}
		\langle Q \rangle_{c} &= \langle H \rangle_{A} - \langle H \rangle_{D} \nonumber \\
		&= \frac{\omega_{c}}{2} \left[\coth \left(\frac{\beta_{c} \omega_{c}}{2}\right) - \lambda_{CD} \coth \left(\frac{\beta_{h} \omega_{h}}{2}\right)\cosh(2r)\right]. \label{eq:6}
	\end{align}
	In the sign convention used here, heat and work added (extracted) to (from) the working fluid are taken to be positive (negative).  Then, from the   first law of thermodynamics, the extracted work from the engine is given by $\langle W\rangle_{\rm ext} = \langle Q\rangle_h+ \langle{Q\rangle_c}$.

	\section{Asymmetric Quantum Otto Heat Engine}
	In the asymmetric Otto engine (AOE), asymmetry is introduced into one of the two work strokes, based on which we will discuss sudden expansion stroke and sudden compression stroke cases.  In both cases, the efficiency of an engine is defined as the ratio of extracted work per cycle to the heat absorbed by the system from a hot reservoir, and is given by 
	\begin{equation}
		\eta = \dfrac{\langle W \rangle_{ext}}{\langle Q \rangle_{h} } .
	\end{equation}
	
	\subsection{Sudden Expansion Stroke}
	
\begin{figure}
		\includegraphics[width=8.8cm, height=7.2cm]{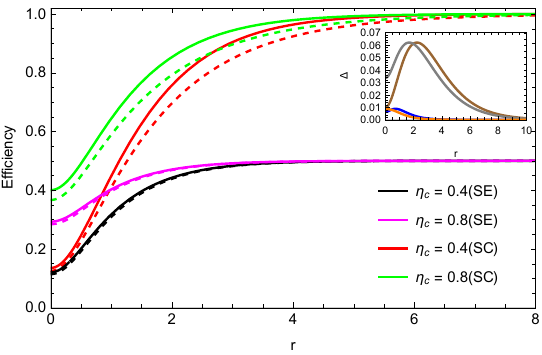}
		\caption{Variation in upper bound on efficiency and efficiency at maximum work as a function of squeezing parameter $r$ for some fixed values of Carnot efficiency in sudden expansion (SE) and sudden compression (SC) stroke cases. Solid black and pink curves represent $\eta_{SE}^{up}$ (Eq.\ref{12}) for $\eta_{c}$ = 0.4 and 0.8, respectively while the solid red and green curves represent $\eta_{SC}^{up}$ (Eq.\ref{eq:9}) for $\eta_{c}$ = 0.4 and 0.8, respectively. Dashed curves in the corresponding colors represent $\eta_{SE}^{MW}$ (Eq.\ref{11}) and $\eta_{SC}^{MW}$ (Eq.\ref{30}). In the inset, we have plotted $\Delta$ = $\eta_{SE}^{up}$ - $\eta_{SE}^{MW}$ (solid blue and orange curves in the inset) and $\Delta^{\prime}$ = $\eta_{SC}^{up}$ - $\eta_{SC}^{MW}$ (solid gray and brown curves in the inset) for $\eta_{c}$ = 0.4 and 0.8 which is always positive means $\eta^{up} > \eta^{MW}$.}
		\label{fig.4}
	\end{figure}

	In this case, expansion stroke (C to D) is considered to be sudden $\left(\lambda_{CD} = (\omega^2_{c}+\omega^2_{h})/2 \omega_{c} \omega_{h}\right)$ while the compression stroke (A to B) is still adiabatic ($\lambda_{AB}$ = 1). By means of Eq.(\ref{eq:5}) and Eq.(\ref{eq:6}), we obtain the work ($W_{SC}$) and efficiency ($\eta_{SC}$) expressions as follows:
	\begin{widetext}
		\begin{equation}
			W_{SE} = \dfrac{(\omega_{h}-\omega_{c})}{4\omega_{h}}\left[\coth\left(\dfrac{\beta_{h}\omega_{h}}{2}\right)\cosh(2r)(\omega_{c}+\omega_{h})-2\omega_{h}\coth\left(\dfrac{\beta_{c}\omega_{c}}{2}\right)\right], \label{01}
		\end{equation}
		\begin{equation}
			\eta_{SE} = \left[\dfrac{2}{1-\dfrac{\omega_{c}^{2}}{\omega_{h}^{2}}}+\dfrac{1}{\left(1+\dfrac{\omega_{c}}{\omega_{h}}\right)\left[\coth\left(\dfrac{\beta_{h}\omega_{h}}{2}\right)\dfrac{\cosh(2r)(\omega_{c}+\omega_{h})}{2\omega_{h}}\tanh\left(\dfrac{\beta_{c}\omega_{c}}{2}\right)-1\right]}\right]^{-1} \equiv \left(\dfrac{2}{f_{1}}+\dfrac{1}{f_{2}}\right)^{-1}.
		\end{equation}
		By positive work condition(PWC), $f_{2}$ $\equiv$	$\left[\coth\left(\dfrac{\beta_{h}\omega_{h}}{2}\right)\dfrac{\cosh(2r)(\omega_{c}+\omega_{h})}{2\omega_{h}}\tanh\left(\dfrac{\beta_{c}\omega_{c}}{2}\right)-1\right]$ $>$ 0 or 1/$f_{2}$ $>$ 0 .
	\end{widetext}	
	Here, $f_{1}$ = 1 - $\omega_{c}^{2}/\omega_{h}^{2}$. As $\omega_{h}$ and $\omega_{c}$ are positive and $\omega_{h}$ $>$ $\omega_{c}$, $f_{1}$ will always lies in the range (0,1). So,
	
	\begin{equation}
		\eta_{SE} < \dfrac{f_{1}}{2} \quad and \quad \eta_{SE} < f_{2} . 
	\end{equation}
	As 0 $<$ $f_{1}$ $<$ 1 and 1/$f_{2}$ $>$ 0, we get:
	\begin{equation}
		\eta_{SE} < \dfrac{1}{2} .
	\end{equation}
	We obtain that the efficiency for sudden expansion stroke is at most one-half, even for a large amount of squeezing. Fig.(\ref{fig.4}) also reconfirms this result by showing that the maximum attainable efficiency in the sudden expansion case is only half. We can explain this efficiency limit as follows: in the sudden expansion stroke, due to the sudden change in the oscillator's frequency, the system can not remain in its instantaneous eigenstate, leading to the nonadiabatic transitions among the energy levels and the system transits into a nonequilibrium state. The state, which was initially diagonal in energy eigenbasis, deviates from its diagonal form due to the presence of off-diagonal terms (popularly known as coherences). The extra cost of creating these coherences is stored in the system as parasitic energy. This excess energy is dissipated at the heat reservoirs and the phenomenon is   referred to as quantum friction \cite{rezek2010reflections,plastina2014irreversible,rezek2010reflections,ccakmak2017irreversible,kosloff2017quantum,latune2021roles,alecce2015quantum,ccakmak2016irreversibility}.

   In Ref. \cite{singh2023asymmetric}, it was shown that in the presence of frictional effects, low-temperature limit is not a appropriate regime for the optimal operation of the engine. We will work in the high-temperature regime as it is advantageous for acquiring analytic results in closed form.  We obtain expressions for the work and efficiency by setting $\coth (\beta_{i}\omega_{i}/2) \approx  2/\beta_{i}\omega_{i}$ ($i = c, h$) :
	\begin{equation}
		W_{SE}^{HT} = \dfrac{(1-z)}{2z\beta_{h}}\left[z(1+z)\cosh(2r)-2\tau\right],\label{eq:22}
	\end{equation}
	\begin{equation}
		\eta^{HT}_{SE} = \dfrac{(1-z)\left[z(1+z)\cosh(2r)-2\tau\right]}{2\left[z\cosh(2r)-\tau\right]} ,\label{eq:13}
	\end{equation}
	where $z =\omega_{c}$/$\omega_{h}$ is he compression ratio   and $\tau$ = $\beta_{h}$/$\beta_{c}$.
	The upper bound on efficiency can be obtained by setting $\partial$$\eta_{SE}^{HT}$/$\partial z$ = 0, which results in the following cubic equation,
	\begin{equation}
		2z^{3}\cosh^{2}(2r)-3z^{2}\tau\cosh(2r)+\left[2\tau-\cosh(2r)\right]\tau=0.
	\end{equation}
	
	As in the case of casus irreducibilis (see Appendix A), the above equation cannot be solved in terms of real radicals. However, we can solve the above equation in terms of trigonometric functions (\ref{appendixB}), and get the following solution\cite{singh2020optimal,singh2020}
	
	\begin{widetext}
		\begin{equation}
			z^{*} = \dfrac{1}{2}\tau \sech(2r)+\cos\left[\dfrac{1}{3}\cos^{-1}\left(\dfrac{2\cosh(2r)+\tau(\tau \sech(2r)-4)}{\tau^{2}\sech(2r)}\right)\right]\tau \sech(2r).\label{2}
		\end{equation}
	\end{widetext}	
	Now substitute Eq.(\ref{2}) in Eq.(\ref{eq:13}), and after simplification, we come up with the following expression of upper bound on efficiency (maximum efficiency) for sudden expansion stroke( $\eta_{SE}^{up}$),
	
	\begin{widetext}
		\begin{equation}
			\eta_{SE}^{up} = \dfrac{\left[\dfrac{1}{4}\sech(2r)\left[1+2\cosh(2r)(1+A)-\eta_{c}\right]\left(2A\cosh(2r)+1-\eta_{c}\right)-2(1-\eta_{c})\right]\left[\sech(2r)+2(A-1)-\eta_{c}\sech(2r)\right]}{2\left[1-\eta_{c}-2A\cosh(2r)\right]} ,\label{12}
		\end{equation}
		where A = $\cos\left[\dfrac{1}{3}\cos^{-1}\left(\dfrac{2\cosh(2r)+\sech(2r)(1-\eta_{c})\left[1-\eta_{c}-4\cosh(2r)\right]}{(1-\eta_{c})^{2}\sech(2r)}\right)\right](1-\eta_{c})\sech(2r)$.
	\end{widetext}
	
	From Eq.(\ref{12}), it is observed that the upper bound on efficiency for sudden expansion stroke depends on the squeezed parameter $r$ and on the ratio of inverse temperatures (or $\eta_{c}$) and is independent of the system's parameters.
	
	To determine the optimal performance of the engine, we calculate the efficiency at maximum work by optimizing Eq.(\ref{eq:22}) with respect to $z$ which results into $z^{*}$ = $\left[\tau \sech(2r)\right]^{1/3}$. Now, by making use of Eq.(\ref{eq:13}), we obtain the following expression of efficiency at maximum work as:
	\begin{equation}
		\eta_{SE}^{MW} = \dfrac{\cosh(2r)(1-2K^2)+(1-\eta_{c})(3-2K)}{2(1+\cosh(2r)-\eta_{c})} .\label{11}
	\end{equation}
	where $K = \left[(1-\eta_{c})/\cosh(2r)\right]^{1/3}$. From Eq.(\ref{11}), it is identified that the efficiency at maximum work depends only on Carnot efficiency and the squeezing parameter $r$ and is independent of the system's parameters. 
	
	For sudden expansion stroke, we plot the efficiency at maximum work output and upper bound on efficiency with squeezing parameter $r$ in Fig.(\ref{fig.4}) and see that the efficiency approaches one-half even for a very large value of squeezing ($r\gg1$). In the inset, we have plotted the difference between $\eta_{SE}^{up}$ and 
	$\eta_{SE}^{MW}$  and see that although  $\eta_{SE}^{up }> \eta_{SE}^{MW}$,  the difference is very small.

	
	\subsection{Sudden Compression Stroke}
	In this case, we take the compression stroke (A to B) as sudden and the expansion stroke (C to D) as adiabatic. The adiabaticity parameters will be $\lambda_{AB} = (\omega^2_{c}+\omega^2_{h})/2 \omega_{c} \omega_{h}$ and $\lambda_{CD}$ = 1. Proceeding as the earlier case,   expressions of work ($W_{SC}$) and efficiency ($\eta_{SC}$) in a high-temperature regime for sudden compression stroke are   obtained as:
	
	\begin{equation}
		W_{SC}^{HT} = \dfrac{(1-z)}{2z^{2}\beta_{h}}\left[2z^{2}\cosh(2r)-\tau(1+z)\right],\label{eq:21}
	\end{equation}
	\begin{equation}
		\eta^{HT}_{SC} = \dfrac{(1-z)\left[2z^{2}\cosh(2r)-\tau(z+1)\right]}{2 z^{2}\cosh(2r)-\tau(z^{2}+1)}.\label{eq:8}
	\end{equation}
	The upper bound on efficiency in sudden compression stroke can be obtained by optimizing Eq.(\ref{eq:8}) with respect to the compression ratio $z$, i.e., by setting $\partial$$\eta_{SC}^{HT}$/$\partial z$ = 0 and we get
	\begin{equation}
		z^{3}\left[2\cosh(2r)-\tau\right]\cosh(2r)-3z\tau\cosh(2r)+2\tau^{2}=0.
	\end{equation}
	As the above equation is again being a case of casus irreducibilis, so the solution can be obtained in terms of trigonometric functions as(see Appendix \ref{appendixA}):
	\begin{equation}
		\small	z^{*} = 2\cos\left[\dfrac{1}{3}\cos^{-1}\left(\dfrac{-\tau \sech(2r)}{\sqrt{\dfrac{\tau}{2\cosh(2r)-\tau}}}\right)\right]\sqrt{\dfrac{\tau}{2\cosh(2r)-\tau}} .\label{1}
	\end{equation}
	Insert Eq.(\ref{1}) in Eq.(\ref{eq:8}) and we will obtain the upper bound on the efficiency for the sudden compression stroke as follows:
	\begin{widetext}
		\begin{equation}
			\small	\eta^{up}_{SC} =\dfrac{\left[1-2 \cos\left(\dfrac{1}{3}B\right)P\right]\bigg\{-1-2 \cos\left(\dfrac{1}{3}B\right)P+\eta_{c}\left[1+2\cos\left(\dfrac{1}{3}B\right)P\right]+\cosh(2r)\left[-2-4 \cos\left(\dfrac{2}{3}B\right)+4 \cos\left(\dfrac{1}{3}B\right)P\right]\bigg\}}{\left[1+2\cos\left(\dfrac{2}{3}B\right)\right]\left[1-\eta_{c}-2\cosh(2r)\right]}\label{eq:9},
		\end{equation}
	\end{widetext}
	where P = $\sqrt{(1-\eta_{c})/(-1+2\cosh(2r)+\eta_{c})}$ and B = $\cos^{-1}\left(-[(1-\eta_{c})\sech(2r)]/P\right)$. In order to obtain the efficiency at maximum work output, we optimize Eq.(\ref{eq:21}) with respect to $z$ and obtain $z^{*}$ = $\left[\tau \sech(2r)\right]^{1/3}$. By using this value in Eq.(\ref{eq:8}), we obtain the following efficiency expression,
		\begin{equation}
			\eta_{SC}^{MW} = \dfrac{(1-K)\left[2K^{2}\cosh(2r)-(1+K)(1-\eta_{c})\right]}{2K^{2}\cosh(2r)-(1+K^{2})(1-\eta_{c})}, \label{30}
	\end{equation}
	where K = $\left[(1-\eta_{c})/\cosh(2r)\right]^{1/3}$. For the sudden compression stroke, we have plotted $\eta_{SC}^{up}$ (Eq.(\ref{eq:9})) and $\eta_{SC}^{MW}$ (Eq.(\ref{30})) with squeezed parameter $r$ for some fixed values of Carnot efficiency in Fig.(\ref{fig.4}). Within the inset, we have graphed the difference between $\eta_{SC}^{up}$ and $\eta_{SC}^{MW}$ and analyze that $\eta^{up} > \eta^{MW}$ for both sudden expansion and compression cases. We reconfirm that the maximum attainable efficiency for sudden compression stroke has a unit value while for sudden expansion, it's only one-half.
	\begin{figure} [H]
		\includegraphics[width=8.8cm, height=6.2cm]{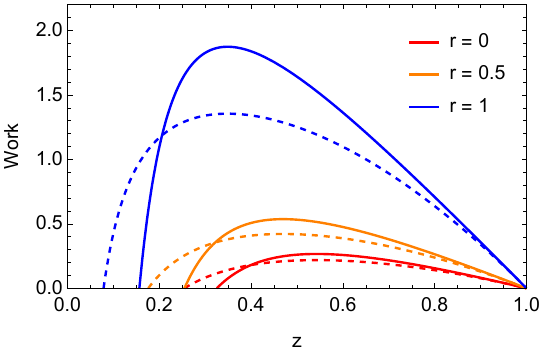}
		\caption{Work output (Eq.\ref{eq:22} and Eq.\ref{eq:21}) as a function of the frequency ratio for different values of squeezing parameter $r$ and fixed value of $\tau$ = 0.16. The solid red, orange, and blue curves are for $r$ = \{0, 0.5, 1\} for sudden compression stroke. Meanwhile, dashed curves in the corresponding colors are for sudden expansion stroke for the corresponding $r$ values. The intersection between the solid and dashed curves is given by $z = \sqrt{\tau \sech(2r)}$ while the maximum work output for both the cases will occur at the same $z$ value given by $z^{*} = \left[\tau \sech(2r)\right]^{1/3}$.}
		\label{fig.9}
	\end{figure}

	 Now, in order to investigate the effect of squeezing on the performance of the heat engine in both configurations, we will plot graphs of the work output and efficiency of the engine as a function of compression ratio $z$ for three different fixed values of squeezing parameter $r$. In Fig.(\ref{fig.9}), we plot  work output as a function of $z$ for different $r$ values with a fixed value of $\tau$ = 0.16. Solid and dashed curves are for sudden compression and sudden expansion strokes, respectively. In both cases, it is clear that the extracted work increases with the increasing value squeezing parameter $r$, thus illustrating the advantages of squeezing. 
	
	In order to compare the performance of the engine between two different cycle configurations,  we obtain the intersection points of the curves by putting $W_{SE}^{HT}$ = $W_{SC}^{HT}$ (from Eqs.(\ref{eq:22}) and (\ref{eq:21})) which results into $z = \sqrt{\tau \sech(2r)}$. 
	Further, the compression ratio $z$, for which the Otto cycle with sudden expansion stroke starts to work as the heat engine, can be evaluated by solving $W_{SE}^{HT} = [(1-z)(z(1+z)\cosh(2r)-2\tau)]/(2z\beta_{h})=0$, which yields   $z = (\sqrt{1+8\tau / \cosh(2r)}-1)/2$. From Fig.(\ref{fig.9}), we can now conclude that for the range $(\sqrt{1+8\tau / \cosh(2r)}-1)/2 < z < \sqrt{\tau\sech(2r)}$, the work extracted for sudden expansion stroke is more than sudden compression stroke and the converse is true for the range $ \sqrt{\tau\sech(2r)} < z < 1 $. Intersection points of the curves have different values for different amount of squeezing. 
	
	\begin{figure}[H]
		\includegraphics[width=8.8cm, height=6.2cm]{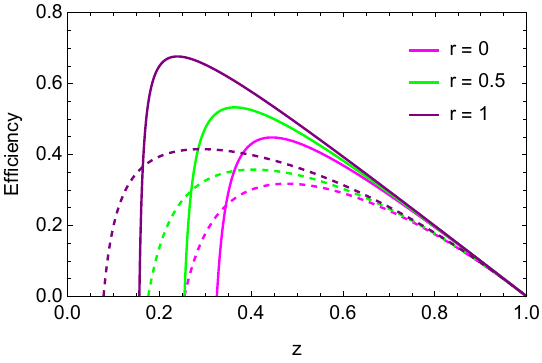}
		\caption{Efficiency for sudden expansion (Eq.\ref{eq:13}) and compression (Eq.\ref{eq:8}) strokes as a function of ratio of frequencies with different values of squeezing parameter $r$ and fixed value of $\tau$ = 0.16. Solid pink, green and purple curves represent efficiency for sudden compression at $r$ = 0, $r$ = 0.5 and $r$ = 1, respectively, while dashed curves in corresponding colors are for sudden expansion stroke for the corresponding values of $r$. The intersection between solid and dashed curves of the efficiency is given by $z = [\tau\cosh(2r)+\sqrt{2\tau\cosh(2r)}(\cosh(2r)-\tau)]/(1+(1-\tau)\cosh(2r)) $.}
		\label{fig.8}
	\end{figure}
	
	Similarly, in Fig.(\ref{fig.8}), we study the efficiency as a function of the frequency ratio $z$ for both the sudden cases. The efficiency plots also show the same trend as that of work output plots. The efficiency plot differs from the work plot only in the intersection point, which here can be derived by putting $\eta_{SE}^{HT}$ = $\eta_{SC}^{HT}$ (from Eqs. (\ref{eq:13}) and (\ref{eq:8})) and results into $z = [\tau\cosh(2r)+\sqrt{2\tau\cosh(2r)}(\cosh(2r)-\tau)]/(1+(1-\tau)\cosh(2r))$ which again depends on the squeezing parameter $r$ and will be different for different curves.

	\section{Full Phase Diagram of the Otto Cycle}
	\begin{table*}[t]
		\centering
		\renewcommand{\arraystretch}{4.5} 
		\caption{Operating conditions and regime of engine, refrigerator, heater and thermal accelerator following the Otto cycle with the squeezing effect for sudden compression and sudden expansion strokes.}
		\label{tab:1}
		\resizebox{\textwidth}{!}{ 
			\begin{tabular}{|c|c|c|c|}
				\hline
				{\textbf{\large Mode}} & {\textbf{\large Condition}} & {\textbf{\large Sudden Expansion Stroke}} & {\textbf{\large Sudden Compression Stroke}}\\ 
				\hline
				{\textbf{\large Engine}} & \large $W_{ext} \geq 0 $, $ Q_{h} \geq 0 $ and $ Q_{c} \leq 0 $ \quad
				& \large $z \geq$ $\dfrac{1}{2}\left(\sqrt{1+\dfrac{8\tau}{\cosh(2r)}}-1\right) $ & \large $z \geq$ $\dfrac{1}{4}\sech(2r)\left[\tau+\sqrt{\tau(\tau+8\cosh(2r))}\right] $ \\
				\hline
				{\textbf{\large Refrigerator}} & \large $W_{ext} \leq 0 $, $ Q_{h} \leq 0 $ and $ Q_{c} \geq 0 $
				& \large $z^{2} \leq \dfrac{2\tau}{\cosh(2r)}-1 $ & \large $ z \leq \dfrac{\tau}{\cosh(2r)} $ \\
				\hline
				{\textbf{\large Heater}} & \large $W_{ext} \leq 0 $, $ Q_{h} \leq 0 $ and $ Q_{c} \leq 0 $
				& \large $ \dfrac{2\tau}{\cosh(2r)}-1 \leq z^{2} \leq \dfrac{\tau^{2}}{\cosh^{2}(2r)} $ & \large $ \dfrac{\tau}{\cosh(2r)} \leq z \leq \sqrt{\dfrac{\tau}{2\cosh(2r)-\tau}} $  \\
				\hline
				{\textbf{\large Thermal Accelerator}} & \large $W_{ext} \leq 0 $, $ Q_{h} \geq 0 $ and $ Q_{c} \leq 0 $
				& \quad \large $ \dfrac{\tau}{\cosh(2r)} \leq z \leq \dfrac{1}{2}\left(\sqrt{1+\dfrac{8\tau}{\cosh(2r)}}-1\right) $ & \large \quad $\sqrt{\dfrac{\tau}{2\cosh(2r)-\tau}} \leq z \leq \dfrac{1}{4}\sech(2r)\left[\tau+\sqrt{\tau(\tau+8\cosh(2r))}\right] $  \\
				\hline
			\end{tabular}
		}
	\end{table*}
	To include the frictional effects, we study the complete phase diagram of the Otto cycle, which gives information about the four operational modes of the cycle, i.e., engine, refrigerator, thermal accelerator, and heater. The latter two come into the picture only by including frictional effects when the processes occur in finite time. For sudden expansion  sudden  compression strokes, we have: 
	\begin{widetext}
		\begin{equation}
			Q^{SE}_{h} = \dfrac{1}{\beta_{h}}\left(\cosh(2r)-\dfrac{\tau}{z}\right), \quad Q^{SE}_{c} = \dfrac{1}{\beta_{h}}\left[\tau-\dfrac{\cosh(2r)}{2}(1+z^{2})\right], 
		\end{equation}
		\begin{equation}
			Q^{SC}_{h} = \dfrac{1}{\beta_{h}}\left[\cosh(2r)-\dfrac{\tau}{2}\left(1+\dfrac{1}{z^{2}}\right)\right], \quad Q^{SC}_{c} = \dfrac{1}{\beta_{h}}\left[\tau-z\cosh(2r)\right].
		\end{equation}
		Along with   Eqs. (\ref{eq:22}) and (\ref{eq:21}), we use the above equations  to find out the different modes of the thermal devices in the high-temperature regime. Details are summarized in Table \ref{tab:1}. Additionally, in Table \ref{tab:1}, we provide the operating regime of these thermal devices under the squeezing  for sudden expansion and sudden compression stroke cases.  
		

		\begin{figure}[H]
			\centering
			\begin{tabular}{ccc}
				\includegraphics[width=0.29\textwidth]{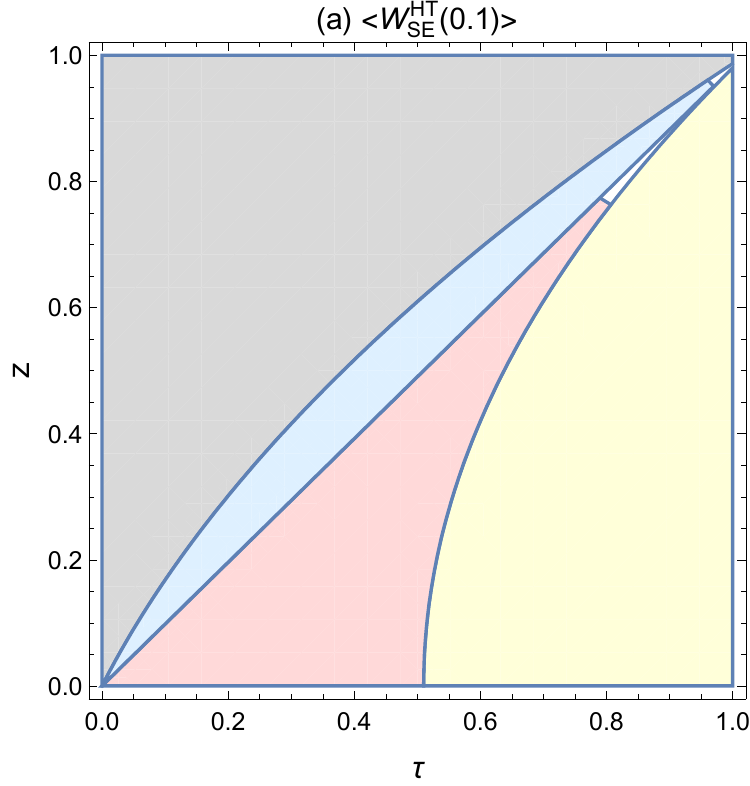} & \quad
				\includegraphics[width=0.29\textwidth]{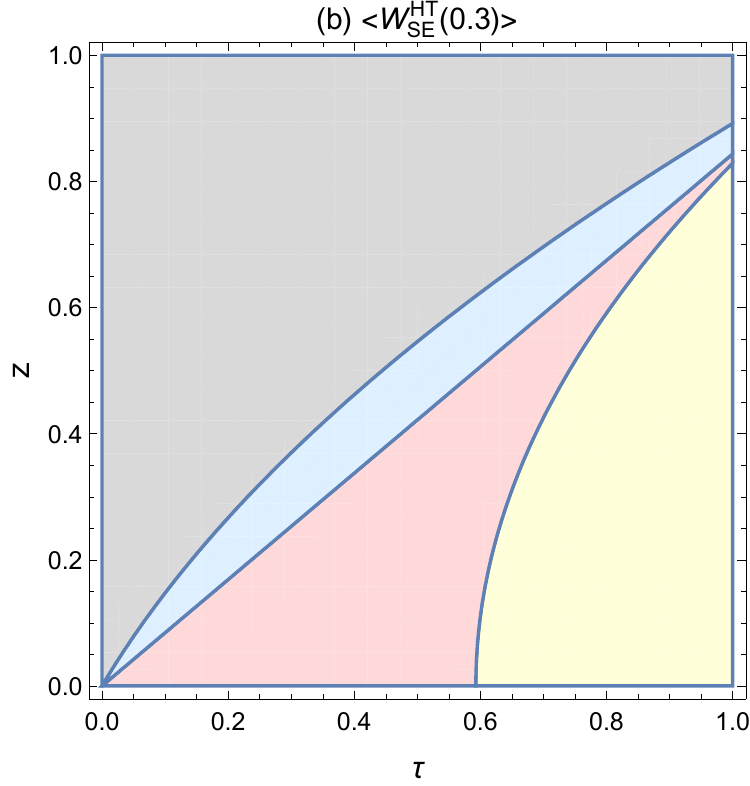} & \quad
				\includegraphics[width=0.373\textwidth,height=5.6cm]{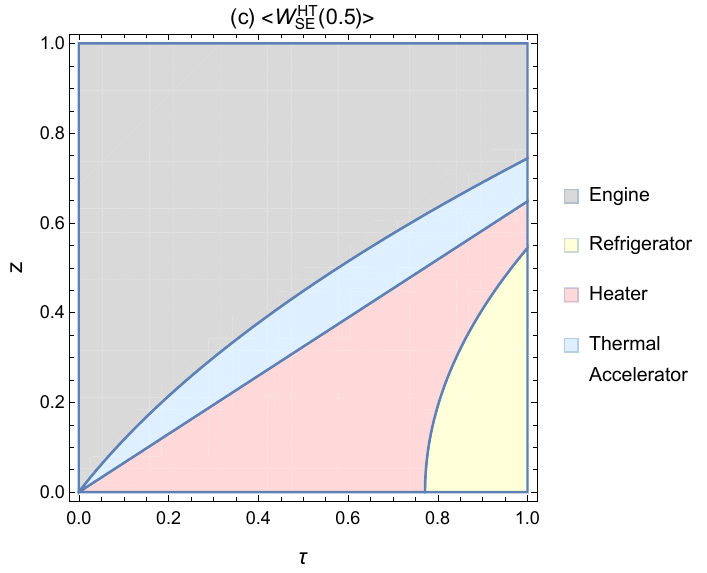} \\
			\end{tabular}
			\begin{minipage}{\textwidth}
				\caption{Complete phase diagram of an Otto cycle as a function of $z = \omega_{c}$/$\omega_{h}$ and $\tau$ = $\beta_{h}$/$\beta_{c}$ for sudden expansion stroke with different squeezing amounts which adopts the values as $r$ = 0.1, $r$ = 0.3, and $r$ = 0.5.}
				\label{fig:8}
			\end{minipage}
		\end{figure}
		\begin{figure}[H]
			\centering
			\begin{tabular}{ccc}
				\includegraphics[width=0.29\textwidth]{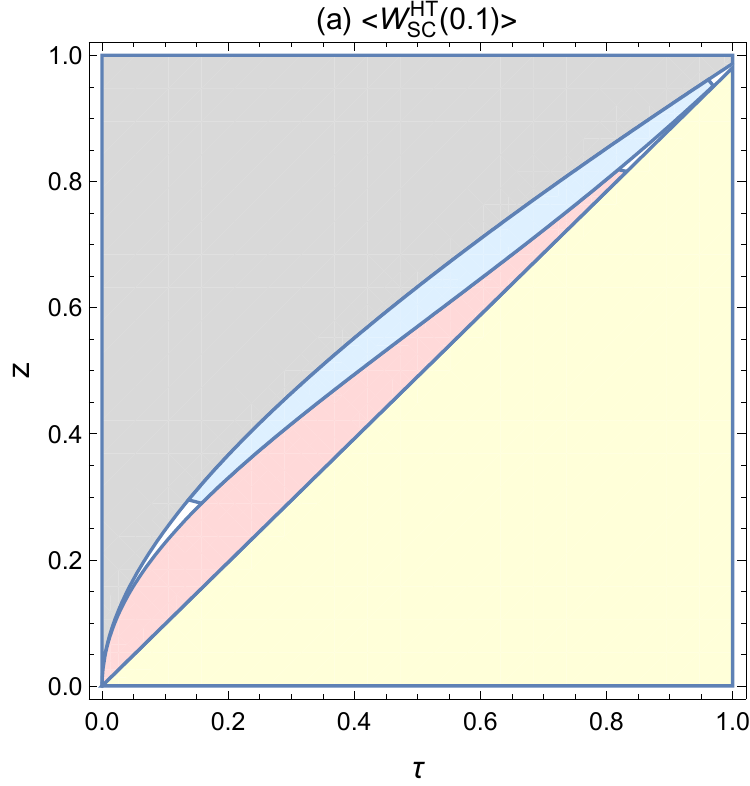} & \quad
				\includegraphics[width=0.29\textwidth]{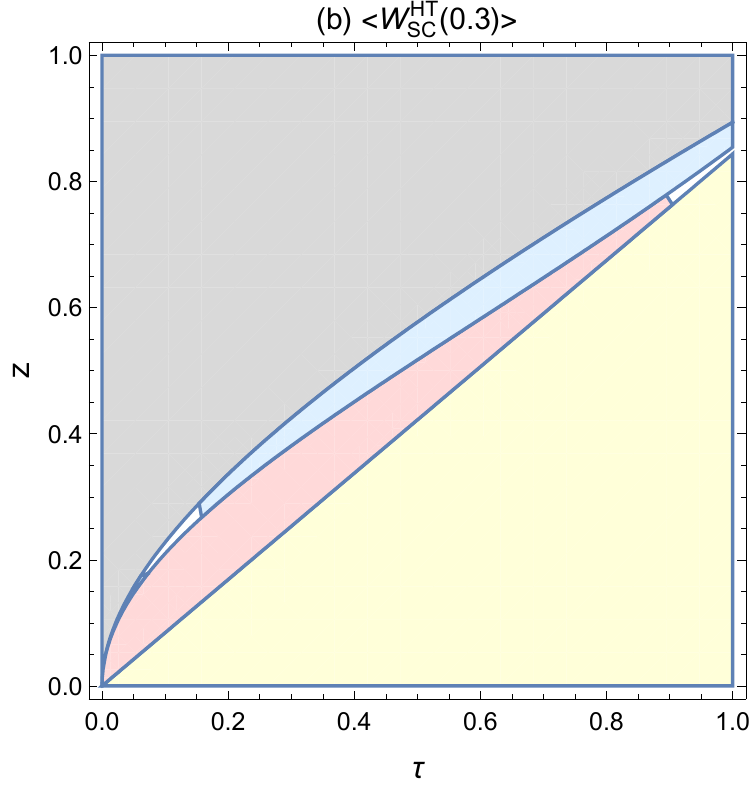} & \quad
				\includegraphics[width=0.373\textwidth,height=5.6cm]{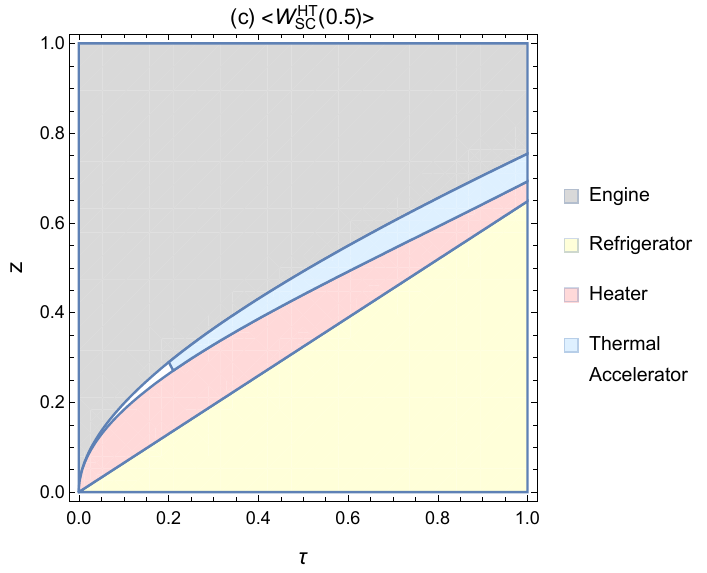} \\
			\end{tabular}
			\begin{minipage}{\textwidth}
				\caption{Complete phase diagram of the Otto cycle as a function of $z = \omega_{c}$/$\omega_{h}$ and $\tau$ = $\beta_{h}$/$\beta_{c}$ with varying $r$ values, i.e., $r$ = 0.1, 0.3, and 0.5 for sudden compression stroke.}
				\label{fig:9}
			\end{minipage}
		\end{figure}
	\end{widetext}
	From both Fig.(\ref{fig:8}) and (\ref{fig:9}), it is observed that for both sudden expansion and compression stroke cases, the parameter regime for the the engine operation increases with increasing degree of squeezing. At the same time, the refrigeration regime goes on decreasing with increasing squeezing for both the cases. For the given values of squeezing parameter $r$ discussed here, the parameter regime for the heater and accelerator modes does not shrink much with increasing squeezing. However, for very large squeezing ($r\rightarrow\infty$), Otto cycle has only engine mode.

	\section{conclusion}
	
	In the present paper, we have studied an asymmetric quantum harmonic Otto engine  with a time-dependent harmonic oscillator as the working material. We have coupled the system with a hot squeezed thermal reservoir and   cold thermal reservoir. By introducing the asymmetry between the work strokes of the asymmetric Otto engine, we have discussed sudden expansion and sudden compression stroke cases. For both cases, we have obtained analytical expressions of the upper bound on efficiency as well as expressions for the efficiency at maximum work.    Then, we studied the effect of the degree of squeezing on the efficiency and work extraction, and  found that the both work output and efficiency  increases with increasing squeezing.

	We have determined a particular range where the engine in sudden expansion configuration performs better than in sudden compression configuration and vice versa. Finally, through the full phase diagrams   (Figs. 5 and 6)  of the Otto cycle for both the configurations under consideration, we reveal that with an increasing squeezing degree, the operational regime of the engine mode grows at the expense of the refrigerator regime.

	\section{Acknowledgment}
	Monika is thankful to the government of India for providing financial support for this work via an Institute fellowship under Dr. B. R. Ambedkar National Institute of Technology Jalandhar. 
	V. S. acknowledges the financial support through the KIAS Individual Grant No. PG096801  at Korea Institute for Advanced Study.

	\appendix
	\begin{widetext}
		\section{Casus irreducibilis}
		
		While solving the cubic equations, there may arise the case of casus irreducibis when the discriminant $D = 18abcd-4b^{3}d+b^{2}c^{2}-4ac^{3}-27a^{2}d^{2}$, of the cubic equation \cite{barnett2002methods}
		
		\begin{equation}
			ay^{3}+by^{2}+cy+d = 0,
		\end{equation}
		is greater than 0, i. e. $D>0$. The above equation can be expressed in the following form,
		
		\begin{equation}
			y^{3}+Ay^{2}+By+C = 0, 
		\end{equation}
		where    $A = b/a$, $B = c/a$ and $C = d/a$.
		The solution of the above equation can be obtained in terms of trigonometric functions and is given by
		
		\begin{equation}
			y = -\dfrac{A}{3}+\dfrac{2}{3}\sqrt{A^{2}-3B}\cos\left[\dfrac{1}{3}\cos^{-1}\left(-\dfrac{2A^{3}-9AB+27C}{2(A^{2}-3B)^{3/2}}\right)\right].
		\end{equation}
		
		We will obtain solution of cubic equations in this way for two cases which are discussed in the present paper.
		\subsection{Sudden Expansion Case}
		\label{appendixB}
		For the sudden expansion stroke, discriminant $D$ of the equation,
		\begin{equation}
			z^{3}-\dfrac{3z^{2}\tau}{2\cosh(2r)}+\dfrac{\tau\left[2\tau-\cosh(2r)\right]}{2\cosh^{2}(2r)} = 0, 
		\end{equation}
		is given by	
		$D =  108\tau^{2}\left[2\tau-\cosh(2r)\right]\left[\tau-\cosh(2r)\right]^{2}\cosh^{3}(2r)>0$ . Here,   $A = -3 \tau/2\cosh(2r)$, $B = 0$, 
		$C = \tau\left[2\tau-\cosh(2r)\right]/2\cosh^{2}(2r)$.

		\subsection{Sudden Compression Case}
		\label{appendixA}
		In our case,   the discriminant of cubic equation
		\begin{equation}
			z^{3}-\dfrac{z\left[3\tau\cosh(2r)\right]}{\cosh(2r)\left[2\cosh(2r)-\tau\right]}+\dfrac{2\tau^{2}}{\cosh(2r)\left[2\cosh(2r)-\tau\right]} = 0
		\end{equation}
		
		will be D = $ 108\tau^{3}\left[2\cosh(2r)-\tau\right]\left[\tau-\cosh(2r)\right]^{2}\cosh^{2}(2r)$$>$0. Here, $ A = 0$, $B = -3\tau\cosh(2r)/\cosh(2r)\left[2\cosh(2r)-\tau\right]$, $C = 2\tau^{2}/\cosh(2r)\left[2\cosh(2r)-\tau\right]$.
		
	\end{widetext}

	\bibliographystyle{apsrev4-2}
	\bibliography{AOE_Squeezing_ref.bib}

\end{document}